\documentclass[aps,prb,twocolumn,groupedaddress,floatfix]{revtex4-2}

\usepackage{graphicx}
\usepackage{dcolumn}
\usepackage{bm}
\usepackage{mathtools}
\usepackage{xcolor}

\usepackage{physics}

\newlength\mylen  \setcitestyle{citesep={,\kern-\mylen}}

\begin{document}

\title{First-principles calculations of electrical conductivities of edge-modified graphene nanoribbons: strain effect}
\author{
Sanjay Prabhakar$^1$ and Roderick Melnik$^2$
}
\affiliation{
$^1$Department of Natural Sciences, Northwest Missouri State University, 800 University Drive, Maryville, MO 64468\\
$^2$MS2Discovery Interdisciplinary Research Institute, M2Net Lab, Wilfrid Laurier University, 75 University Avenue, Waterloo, ON N3L 3V6, Canada
}

\date{March 26, 2022}

\begin{abstract}
We investigate the influence of strain on the electrical properties of graphene nanoribbons that have potential applications in making sensors and other optoelectronic devices. In particular, we chose pristine armchair graphene nanoribbons with 7 zigzag edges (7aGNRsH), boron doped  armchair graphene nanoribbons with 7 zigzag edges (7aGNRsH-B) and  armchair graphene nanoribbons with 7 zigzag edges that have one carbon atom vacancy (7aGNRsH-V). Based on first-principles calculations, results show that pristine unstrained 7aGNRsH is electrically  nonconductive but turns to be electrically conductive in a wide range of energy spectrum, e.g., from IR to visible to UV,  due to the application of strain engineering.  In metallic unstrained and strained 7aGNRsH-B and 7aGNRsH-V, non-vanishing electrical conductivity in the IR, visible and UV  energy spectrum regimes are observed. We also investigate the influence of strain on the Berry curvature of 7aGNRsH, 7aGNRsH-B and 7aGNRsH-V nanoribbons. The results show that fermions are spread through out the Brillion zone in the reciprocal space for semiconducting unstrained 7aGNRsH but localized near the $\Gamma$-point for strained  7aGNRsH that have out-of-plane deformations due to strain engineering. For metallics 7aGNRsH-B and 7aGNRsH-V, Berry curvature plots show that fermions are localized far away from the $\Gamma$-point. In two atom boron doped p-type armchair graphene nanoribbons with 7 zigzag edges (7aGNRsH-2B), large peaks in electrical  conductivity at IR energy spectrum regimes can be observed. These peaks of electrical conductivities in 7aGNRSH-2B may be detectable in experimentally synthesized structure in Reference, JACS 137, 8872 (2016).

\end{abstract}

\maketitle
\section{Introduction}
Two dimensional materials, e.g., graphene and others, are of great interest for making optoelectronic and spintronic devices that run at much faster speed compared to conventional silicon and other conventional semiconductor  devices because they possess extraordinary physical properties, e.g., the half integer quantum Hall effect, non-zero Berry curvature, high mobility charge carriers, and other unique properties~\cite{liu20,balents20,liu20a,parker21,geim-nature07,coleman-science11,novoselov05a,novoselov05,novoselov04,savage-nature12}.
In addition, experimental studies have shown that the optoelectronic devices made out of graphene have better performance over  silicon
devices of the same size~\cite{savage-nature12,novoselov04,liao-nature10,aoki13}. Unfortunately, graphene has some drawbacks, e.g., the conduction and valence bands touch each other at the Dirac point. In other words, it  contains no bandgaps and thus it may not have any application at the semiconductor device level. However,  experimentalists have discovered some artificial way to create some bandgaps, e.g.,  strain engineering is one of them and thus can be utilized to make optoelectronic devices. Another methods to open bandgap in graphene is to implement spin-orbit coupling effect, magnetic fields or simply consider bilayer graphene and apply positive and negative voltages on the top and bottom layer~\cite{trauzettel07}. In addition, devices made from armchair graphene nanoribbons possess a larger bandgap opening at $\Gamma$-point~\cite{han-prl07,zhou-nature07,xia-nanoletters10,chen-nature15,ugeda-nature14,brey06}.

\begin{figure*}
\includegraphics[width=17cm,height=4cm]{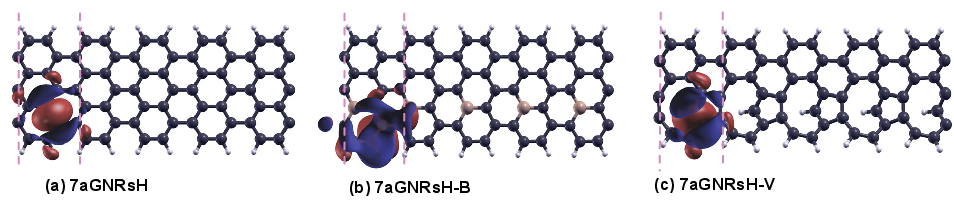}
\caption{\label{fig1HVB} Maximally localized wannier function (MLWF) of top of the valence bands for three different optimized structures: (a) 7aGNRsH (b) boron doped 7aGNRsH and (c) carbon vacancy created 7aGNRsH. Note that in (b) and (c), periodicity of boron atoms and ring-pentagon in the crystal lattice turns electrically  inactive pristine 7aGNRsH from semiconducting into metallic that is electrically active in the wide range of energy spectrum, e.g., IR, visible and UV. For details see conductivity plots in Figs.~(\ref{fig6HVBcond}) and ~(\ref{fig11Kubo}).  }
\end{figure*}

\begin{figure*}
\includegraphics[width=12cm,height=12cm]{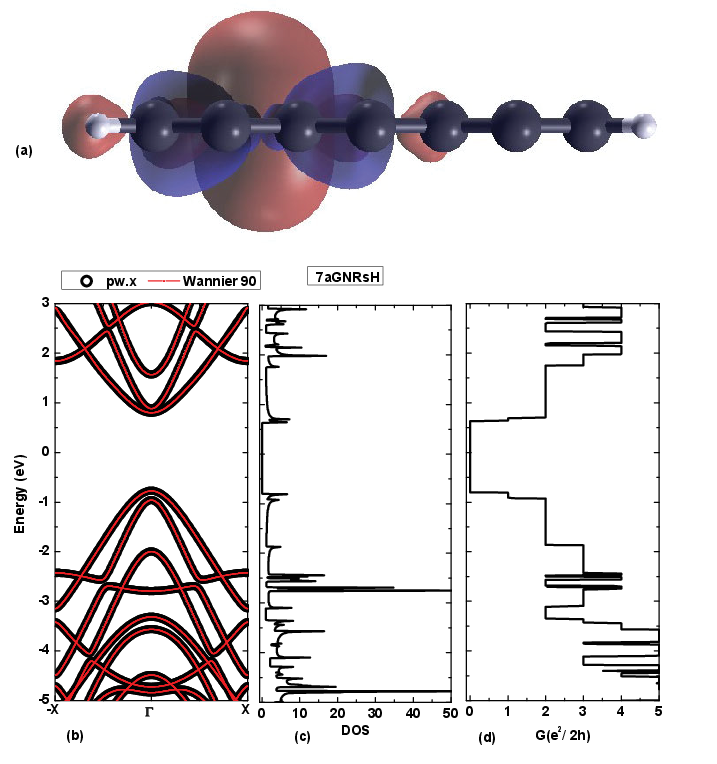}
\caption{\label{fig2bandH} Side view of maximally localized wannier function of top of the valence band states of 7aGNRsH is shown in (a) and  its band structures obtained from first-principles calculations using pw.x (black, open circles) and wanniwer90 (red line with dots) methods are shown in (b). As can be seen, band energies obtained from both methods are in excellent agreement. Also, density of states and quantum conductivity of 7aGNRsH are plotted in (c) and (d). Notice that forbidden bandgap shown in band structure calculation in (a) is also reflected in the density of states and quantum conductivity plots in (c) and (d) respectively.     }
\end{figure*}
\begin{figure*}
\includegraphics[width=12cm,height=16cm]{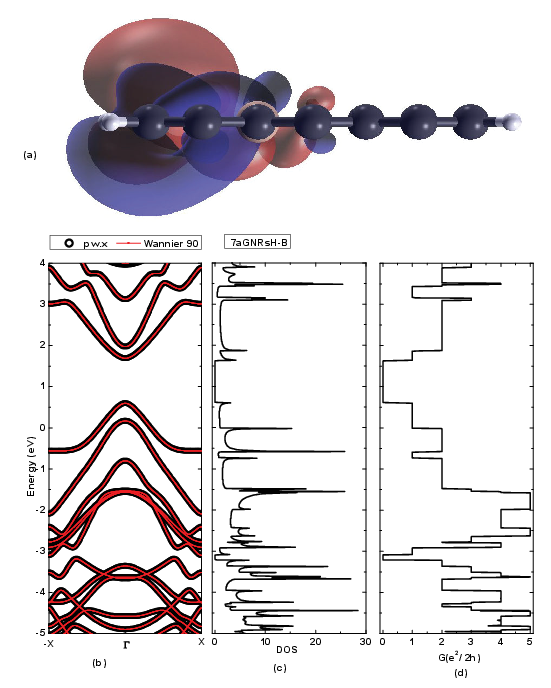}
\caption{\label{fig3bandB} Side view of maximally localized wannier function of top of the valence band states of single boron atom doped 7aGNRsH is shown in (a) and  its band structures obtained from first-principles calculations using pw.x (black, open circles) and wanniwer90 (red line with dots) methods are shown in (b). As can be seen, band energies obtained from both methods are in excellent agreement. Fermi-energy is set to be at 0 eV that is located below the top of the valence band. Also, density of states and quantum conductivity of 7aGNRsH are plotted in (c) and (d). Notice that forbidden bandgap shown in band structure calculation in (a) lies above the Fermi energy that is also reflected in the density of states and quantum conductivity plots in (c) and (d) respectively. Hence, 7aGNRsH-B acts like metallic.   }
\end{figure*}
\begin{figure*}
\includegraphics[width=12cm,height=16cm]{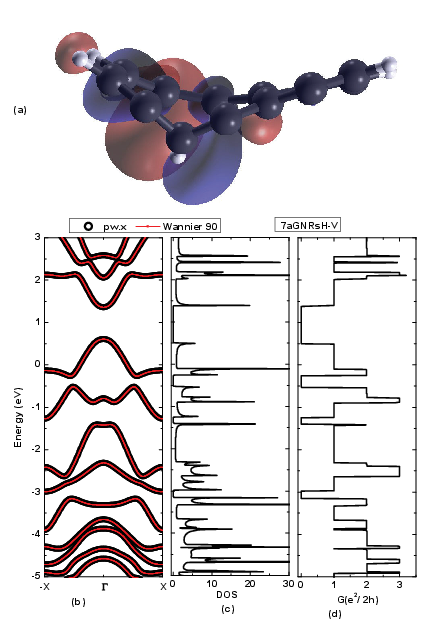}
\caption{\label{fig4bandV} Side view of maximally localized wannier function of top of the valence band states of vacancy of single carbon atom in 7aGNRsH is shown in (a) and  its band structures obtained from first-principles calculations using pw.x (black, open circles) and wanniwer90 (red line with dots) methods are shown in (b). As can be seen, band energies obtained from both methods are in excellent agreement. Fermi-energy is set to be at 0 eV that is located below the top of the valence band. Also, density of states and quantum conductivity of 7aGNRsH are plotted in (c) and (d). Notice that forbidden bandgap shown in band structure calculation in (a) lies above the Fermi energy that is also reflected in the density of states and quantum conductivity plots in (c) and (d) respectively. Hence, 7aGNRsH-V acts like a metallic structure.    }
\end{figure*}
%

The electrical properties of graphene systems have attracted considerable interest for optoelectronic devices, especially to make sensors devices  due to the extension of tunability  of electrical conductivities in the wide range of optical energy spectrum regimes, e.g., from infrared to visible to UV light spectrum~\cite{mishchenko20,wright09,zhang08,stauber08,cserti07,chao-Yang21,hipolito19,herrera20,van16,filippis14}. Recently there have been a lot of research publications on the properties of electrical conductivity of single and bilayer graphene~\cite{gallagher19,narozhny19,li19,iurov18}.   When the fermi level of devices change, then the electrical  transition of the charge carriers within the band energy  ranges from infrared to visible to UV energy spectrum regimes but the absorption coefficient of graphene is still found to have very small, typically less than $3\%$ absorption coefficient that may not attract considerable attention from experimentalists to design devices at the production level~\cite{gusynin06,mojarro21,falomir18,novko16}.

In this paper, we consider three systems: intrinsic armchiar graphene nanoribbons with 7 zigzag edges (7aGNRsH), metallic one boron atom doped armchair graphene nanoribbons with 7 zigzag edges (7aGNRsH-B), metallic  armchair graphene nanoribbons with 7 zigzag edges that have one carbon atom vacancy (7zaGNRsH-V) and p-type 2 boron atoms doped  armchair graphene nanoribbons with 7 zigzag edges (7aGNRsH-2B). We then study the influence of strain on the electrical conductivity of these materials on the wide range of optical energy spectrum. Note that periodicity of boron atom is seen in the optimized structure of 7aGNRsH-B, and periodicity of ring-pentagon is seen in the optimized structure of 7aGNRsH-B. Also, optimized structure of 7aGNRsH-2B has been  exactly the same that is experimentally synthesized in Ref.~\cite{cloke15}. Hence, providing data coming from first-principles calculations for electrical conductivities of site-specific substitutional of two boron doping in semiconducting armchair graphene nanorobbons may help to understand the underlying physics of how charge carries move in a device made out of graphene nanoribbons~\cite{cloke15}.  Furthermore,  first-principles  calculations studies  show  that pristine unstrained 7aGNRsH is electrically not conductive but turns to be conductive in the wide range of optical energy spectra, e.g., from IR to visible to UV with the application of strain engineering. In 7aGNRsH-B, we always find non-vanishing conductivity  due to periodicity of boron doping in the crystal lattice that turns semiconducting pristine material into metallic. Similarly non-vanishing conductivity can also be found in metallic 7aGNRsH-V due to periodicity of ring-pentagon in the crystal lattice. In addition, we also investigate the influence of strain on the Berry curvature of pristine, metallic 7aGNRsH-B, 7aGNRsH-V and p-type 7aGNRsH-2B. The results show that fermions are spread through out the Brillion zone in the reciprocal space for semiconducting 7aGNRsH that have no out-of-plane deformations but localized near the $\Gamma$-points that have out-of-plane deformations due to excessive strain, e.g., about 18$\%$ of compressive strain. For other materials, e.g., metallic 7aGNRsH-B and 7aGNRsH-V, Berry curvature plots show that fermions are localized far away from the $\Gamma$-point. Since, we are interested in investigating the influence of strain on the electrical properties of armchair graphene nanoribbons, there is a variety of ways that allow to design and control the strain in graphene~\cite{prabhakar19prb}. For examples, in-plane strain can be induced by applying compressive stress throughv the armchair edge and tensile stress through the zigzag edge in graphene nanoribbons. Out-of-plane deformations can also be observed for sufficiently large compressive stress, (strain should exceed approximately $18\%$ in pristine seven armchair graphene nanoribbons) through the armchair edge. Out-of-plane deformations mostly observed due to adsorption of additives (e.g., H or OH) on the surface of the graphene nanoribbons~\cite{deepika15,lim15, bao-nature09,prabhakar16,christensen-prb15,prabhakar14,deepika15,lim15,lee08,khestanova16,ding10}.
One may also induce strain by placing graphene nanoribbons on the  substrate (e.g. SiC) that can induce significantly large amount of strain due to lattice mismatch between graphene nanoribbons and SiC substrate~\cite{ni08,bastos16}.
Strain engineering of ripples and wrinkles is of greate interest for straintronic  applications~\cite{bronsgeest15,cerda03,ryan17,prabhakar14,prabhakar16,ferralis08,boyd15}.
Recent experimental studies on band engineering tuning due to strain engineering in graphene and TMD 2D materials are discussed in Ref.~\cite{peng20strain}. Strain engineering of ripples and wrinkles is of great interest for straintronic  applications~\cite{bronsgeest15,cerda03,ryan17,prabhakar14,prabhakar16,ferralis08,boyd15}.
Several works suggest that graphene can sustain very large elastic strains (up to 25$\%$), comparable to Griffith’s
theoretical limit~\cite{griffith1921vi,cocco10,naumov11}.


The paper is organized as follows. In Sec.~\ref{theoretical-model}, we provide computational details of first-principles calculations where we use the Quantum Espresso software package and the wannier90 program~\cite{giannozzi09,pizzi20}. In section~\ref{results-discussion}, we present the results for calculations of band structures, density of states, quantum conductivity, Berry curvature and electrical conductivity for pristine 7 armchair graphene nanoribbons (7aGNRsH) as well as for metallic 7aGNRsH (7aGNRsH-B, 7aGNRsH-V) and p-type semiconductor 7aGNRsH (7aGNRsH-2B). Finally, in section~\ref{conclusion}, we summarize our results.


\begin{figure}
\includegraphics[width=8.5cm,height=8cm]{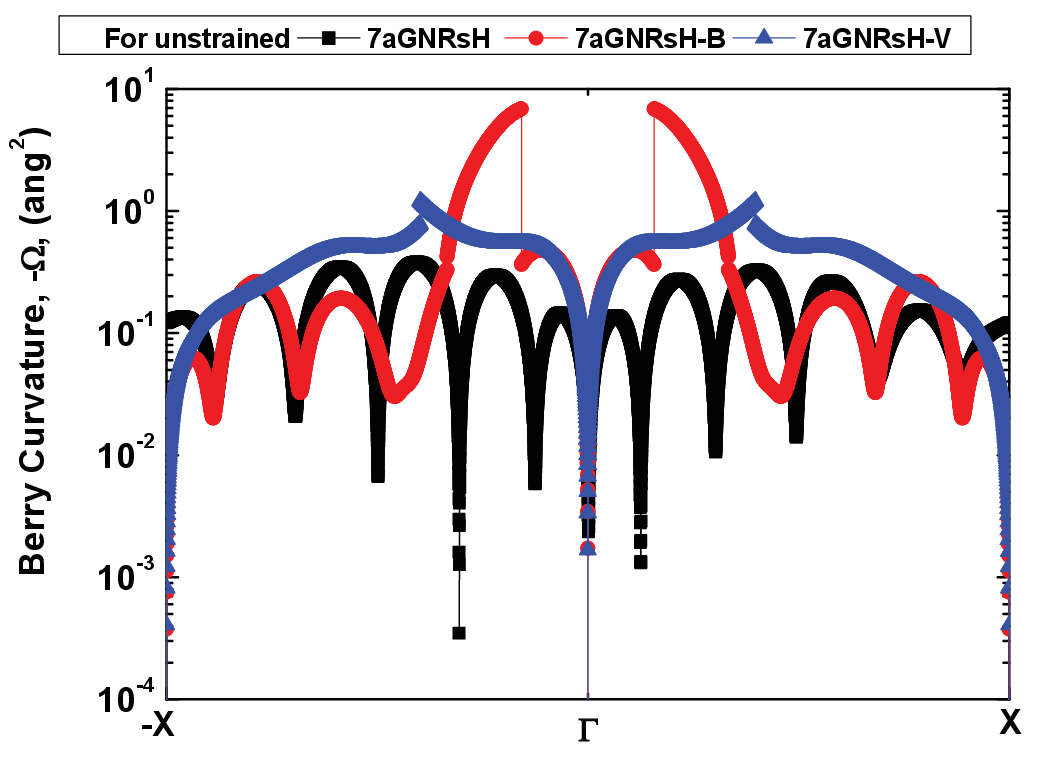}
\caption{\label{fig5BerryCurvature} Berry curvatures of fermions, that are transported from $-X$ to $\Gamma$ to $+X$ in the reciprocal space of the Brillion zone  in an unstrained 7aGNRsH (black line with diamonds), 7aGNRsH-B (red line with circles) and 7aGNRsH-V (blue line with triangles), are plotted. Discontinuity in the Berry curvatures can be seen away from the $\Gamma$-point for 7aGNRsH-B and 7aGNRsH-V due to the presence of Fermi energy below the top of the valence bands and hence 7aGNRsH-B and 7aGNRsH-V are metallic structure.      }
\end{figure}
\begin{figure}
\includegraphics[width=8.5cm,height=8cm]{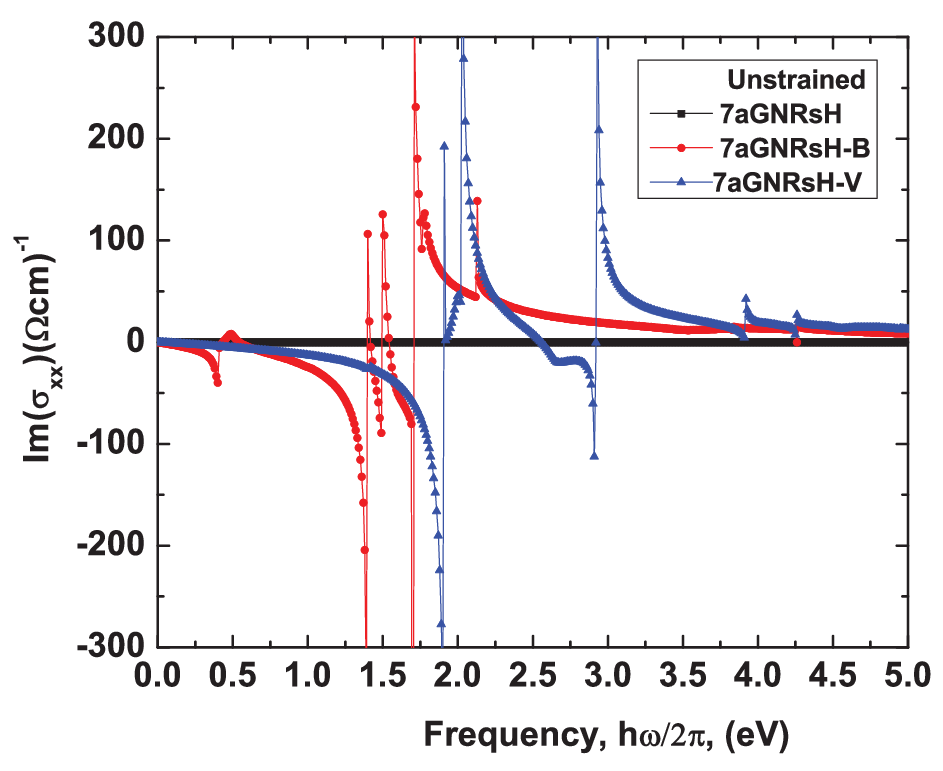}
\caption{\label{fig6HVBcond} Frequency dependence of imaginary part of symmetric conductivity tensor, $\sigma_{xx}$, obtained from Eq.~\ref{kubo-H} by using Kubo-Greenwood formalism. As can be seen, unstrained  pristine 7aGNRsH is electrically inactive but 7aGNRsH-B and 7aGNRsH-V turn to be electrically active in the IR, visible and UV regime of energy spectrum because 7aGNRsH-B and 7aGNRsH-V are metallic due to the presence of Fermi energy below the top of the valence band.     }
\end{figure}
\begin{figure}
\includegraphics[width=8.5cm,height=4cm]{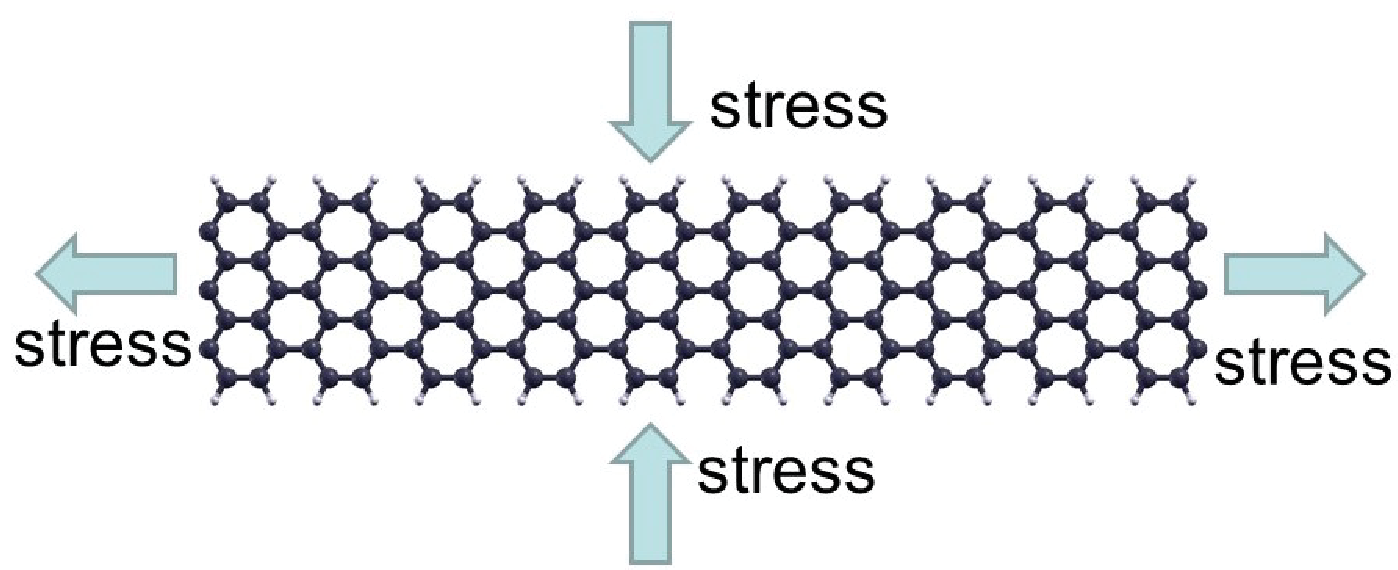}
\caption{\label{fig7Strain} Schematics of armchair graphene nanoribbons with 7 zigzag edges, where tensile stress is applied at the zigzag edge, i.e, along x-direction  and compressive stress is applied at the armchair edge, e.g. along y-direction. Such kind of applied stress allows us to investigate the influence of strain on the band engineering, as well as electrical conductivity measurements.
   }
\end{figure}
\begin{figure}
\includegraphics[width=8.5cm,height=8cm]{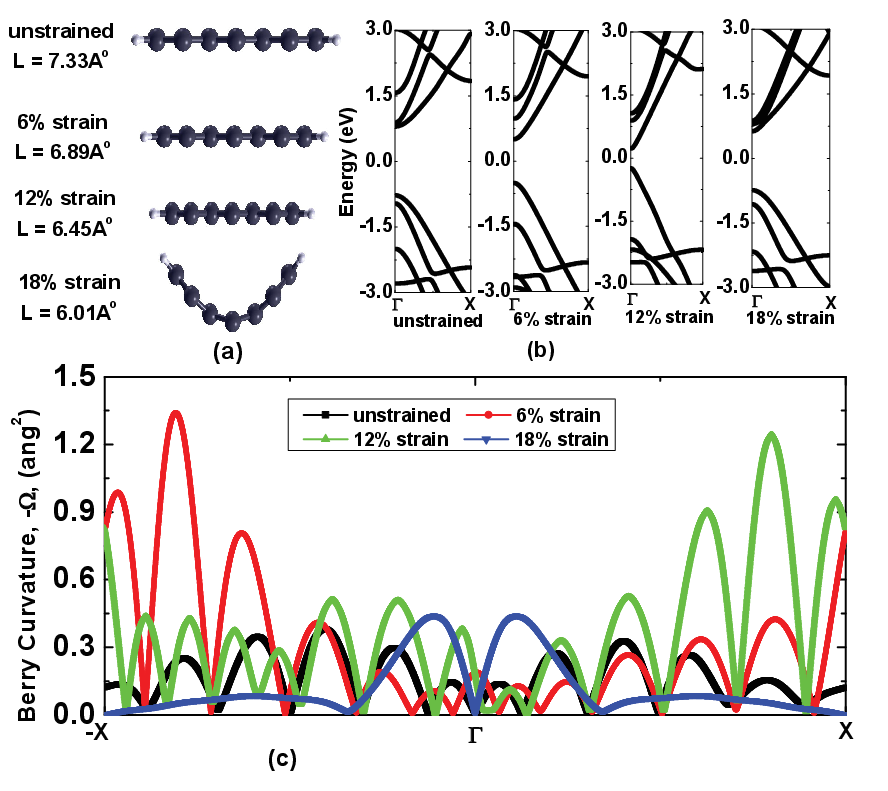}
\caption{\label{fig8Berry} Influence of strain on the width of the 7aGNRsH in (a), band structures in (b) and Berry curvature in (c). As can be seen in the Berry curvature plots, fermions are spread through out the Brillion zones for unstrained armchair graphen nanoribbons but they are localized near gamma point for large strain (approximately 18$\%$ strain) due to relaxed shape graphene nanoribbons, that have out-of-plane deformations (e.g., see Fig\ref{fig8Berry}(a).    }
\end{figure}
\begin{figure}
\includegraphics[width=8.5cm,height=8cm]{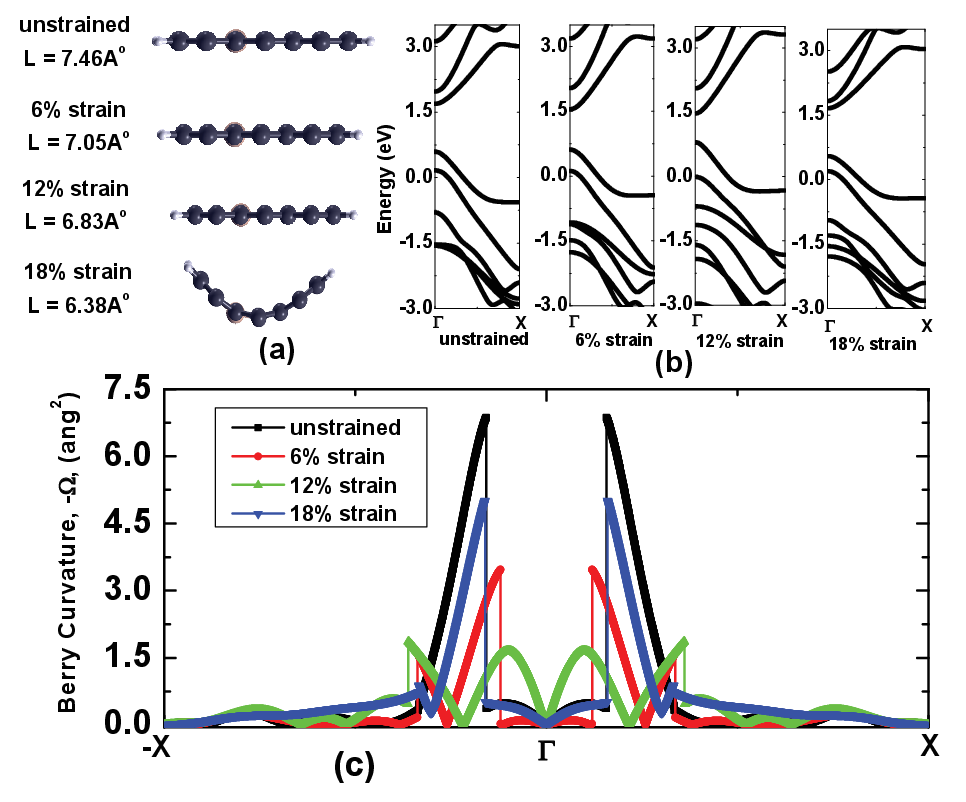}
\caption{\label{fig9Berry} Same as Fig.~\ref{fig8Berry} but for single boron atom doped in 7aGNRsH.    }
\end{figure}
\begin{figure}
\includegraphics[width=8.5cm,height=8cm]{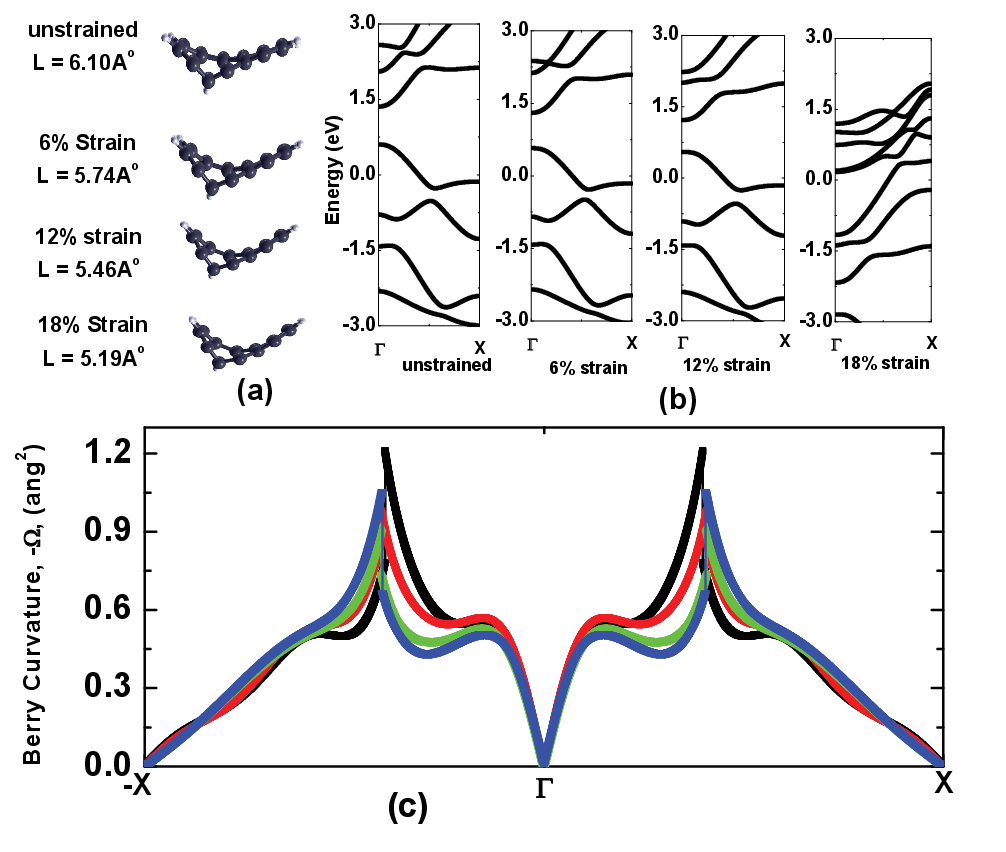}
\caption{\label{fig10Berry} Same as Fig.~\ref{fig8Berry} but for single carbon  atom vacancy  in 7aGNRsH.}
\end{figure}
\begin{figure*}
\includegraphics[width=18cm,height=7cm]{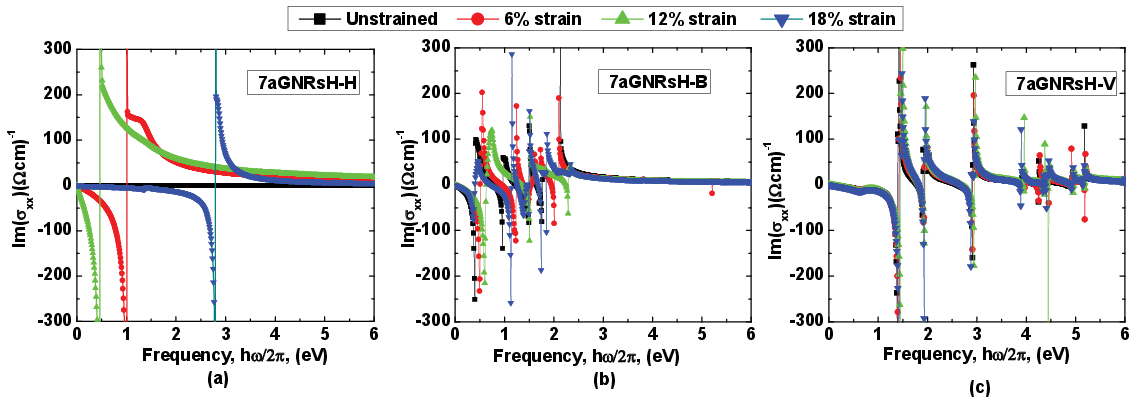}
\caption{\label{fig11Kubo} Influence of strain on the imaginary part of  conductivities, $\sigma_{xx}$, at wide range of energy spectrum regime for intrinsic semiconductor 7aGNRsH in (a), metallic 7aGNRsH-B in (b) and metallic 7aGNRsH-V in (c). Notice that pristine unstrained 7aGNRsH is electrically inactive (black line with diamond in (a)) but turns to be  active in IR spectrum regime for $6\%$ and $12\%$ strained 7aGNRsH (red line with circles and green line with triangles in (a)) and visible spectrum for $18\%$ strain (blue line with triangles pointing down in (a)). Similarly, non-vanishing electrical conductivity in IR and visible energy spectrum regime  are observed in metallic 7aGNRsH-B in (b) in both unstrained and strained cases. Also,  non-vanishing electrical conductivity in IR and visible and UV energy spectrum regime  are observed in metallic 7aGNRsH-V  in both unstrained and strained cases in (c).     }
\end{figure*}
\begin{figure*}
\includegraphics[width=18cm,height=12cm]{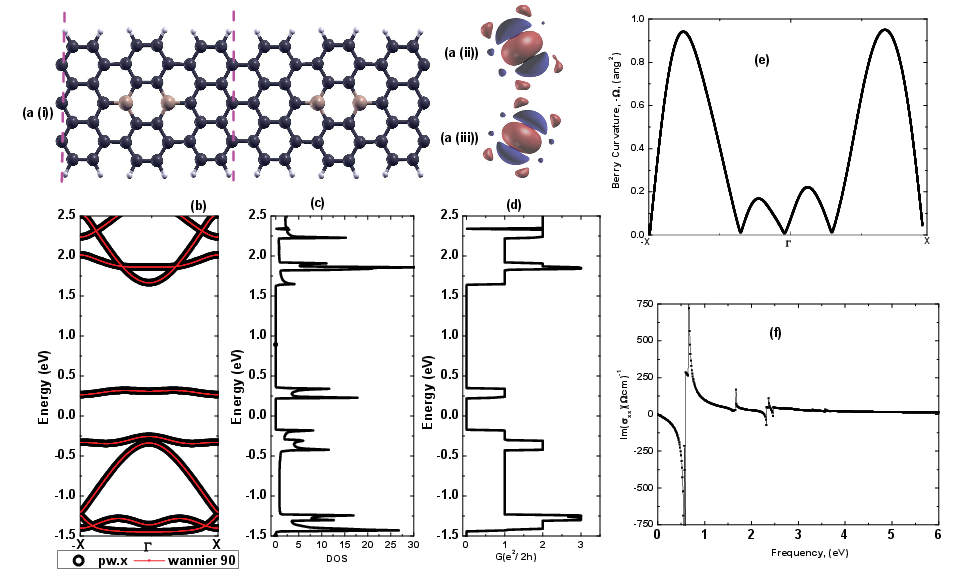}
\caption{\label{fig12-2B-Band} (a) (i) Optimized structure of two boron doped armchair graphene nanoribbons with 7 zigzag edges (7aGNRsH-2B) that was experimentally synthesized in Ref.~\cite{cloke15} and its maximally localized wannier functions of top of the valence and bottom of the conduction bands in (a)(ii,iii). Band structures of 7aGNRsH-2B obtained from pw.x (black open cirlces) and maximally localized wannier functions method (red lines with diamond) are shown in (b). Fermi energy, $\epsilon_F$ is set to be at 0eV. The bandgap between valence band and dopant state is 0.6eV, whereas quasi-particle bandgap between valence and second conduction band is 2.0eV. Such calculated bandgap is also reflected in the density of states and quantum conductivity plots in (c) and (d). The Berry curvature and electrical conductivity plots are shown in figures (b) and (c).   }
\end{figure*}

\section{Computational Methods}\label{theoretical-model}
Density Functional Theory (DFT)  calculations for edge modified armchair graphene nanoribbons with 7 zigzag edges  are performed  in the Quantum Espresso software package~\cite{giannozzi09}, where periodic boundary conditions are implemented.  Ultrasoft pseudopotentials and a plane wave basis set with  kinetic energy and charge density  cut-off at 100Ry and 800Ry are used. We  include exchange and correlation effects within the Perdew-Burke-Ernzerhof (PBE) Functional~\cite{perdew96}. We use orthorhombic  4.34${\AA}$ $\times$ 24.78${\AA}$ $\times$ 15.01${\AA}$ size of supercell that contains an armchair graphene nanoribbons with 7 zigzag edges. In the supercell, we have 18 atoms for the graphene nanoribbons with armchair edge functioning with H (7aGNRsH), 18 atoms for the boron doped graphene nanoribbons  with armchair edge functioning with H (7aGNRsH-B) and 17 atoms for the vacancy created graphene nanoribbons with armchair edge functioning with H (7aGNRsH-V). The optimized lattice parameters along the x-direction are 4.29${\AA}$ for 7aGNRsH, 4.35${\AA}$ for 7aGNRsH-B and 4.37${\AA}$ for 7aGNRsH-V. For the structure synthesized in Ref.~\cite{cloke15}, we have 40 C-atoms, 12 H-atoms and 2 B-atoms in a supercell and  the optimized lattice constant in x-direction is $13.04{\AA}$ that has width of $7.31 {\AA}$. During geometry optimization, all atoms in all  three directions and x-axis of the supercell are fully relaxed until the forces on atoms are smaller than 0.01 eV/$\AA$. We have tested several k-point samplings and calculations are performed at (6,1,3) $k$-points sampling  fulfilling the above convergence criteria. The XCRYSDEN program is used to draw the molecular structure~\cite{kokalj99}. For strain engineering of graphene nanoribbons, we have applied the compressive stress through the armchair edge and  frozen the y-coordinates of the armchair edge atoms while  allowing the other coordinates to be fully relaxed. The x-axis of the supercell is also fully relaxed. For the calculation of Berry curvature, quantum and  electrical conductivities, we have used maximally localized Wannier functions method employed in the Wannier90 program~\cite{pizzi20}.  Atom-center s-like orbital is chosen for the initial projection for H and B atoms as well as at the center of bond length. Also, atom-center p-like orbital is chosen for the initial projection of each C atoms in the lattice. The states below the Fermi energy is frozen during wannier 90 calculations.

\section{Results and Discussions}\label{results-discussion}

Bandgaps of the armchair graphene nanoribbons depend on the width of the ribbons. If the armchair nanoribbons contains an odd number of zigzag edges then the armchair graphene nanoribbons possesses a finite bandgap opening at the $\Gamma$ point and become interesting  for making  semiconductor optoelectronic devices. For an even number of zigzag edges, the armchair graphene nanoribbon has a zero bandgap and thus possesses metallic properties. Bandgaps can be modified either by doping or creating a vacancy or by some other techniques that have application in optoelectronic devices.  In Fig.~\ref{fig1HVB}, we have plotted an optimized structure of armchair graphene nanoribbons with 7 zigzag edges (7aGNRsH), boron doped armchair graphene nanoribbons  with 7 zigzag edges (7aGNRsH-B) and vacancy of one carbon atom in  armchair graphene nanoribbons with 7 zigzag edge(7aGNRsH-V). In addition, we have also plotted the maximally localized wannier functions of the top of the valence band states. As can be seen, wannier functions are well localized within the supercell of the armchair graphene nanoribbons.
Maximally localized wannier functions of the bottom of the conduction bands states that look exactly identical to the top of the valence band states are not shown because the fermi-energy of 7aGNRsH, 7aGNRsH-B and 7aGNRsH-V stay below the top of the conduction band and hence conduction band states have no contribution in electrical conductivities-vs-frequency plots.
For 7aGNRsH-B, one can see the periodicity of B atoms in the crystal lattice and for 7aGNRsH-V, one can find the periodicity of ring-pentagon structures in the crystal lattice that turns these structures from semiconducting into metallic. Comparing the optimized structures of 7aGNRsH (Fig.~\ref{fig1HVB}(a)) with 7aGNRsH-B (Fig.~\ref{fig1HVB} (b)), one finds that three fold rotational symmetry is broken due to Jahn-Teller distortion and a nine sided ring adjacent to the pentagon is formed.

In Fig.~\ref{fig2bandH}(a), we have  plotted side view of the maximally localized wannier function in 7aGNRsH and bandstructures, density of states and quantum conductivity in Fig.~\ref{fig2bandH}(b), (c) and (d) respectively. As can be seen in Fig.~\ref{fig2bandH}(b), the band structures obtained from first-principles calculations using pw.x method~\cite{giannozzi09} are in excellent agreement with the band structures obtained from maximally localized wannier functions method~\cite{pizzi20}. In Fig.~\ref{fig2bandH}(c) and (d), we have used maximally localized wannier functions method and plotted Density of States (DOS) and quantum conductivity of 7aGNRsH. It can be seen that the bandgap of 7aGNRsH is 1.57 eV that is consistent with the values obtained from band structures, density of states and quantum conductivity plots in Figs.~\ref{fig2bandH}(b,c,d). The Fermi energy is set to be at zero eV and it exits at the middle of the forbidden bandgap of 7aGNRsH.

In Fig.~\ref{fig3bandB} (a), we have plotted side view of the optimized structure of one atom boron doped 7aGNRsH (7aGNRsH-B) along with maximally localized wannier functions of the top of the valence band states. For 7aGNRsH-B, the fermi energy, $\epsilon_F = 0$ exits below the top of the valence band states that can be seen in the bandstructures calculation in  Fig.~\ref{fig3bandB}(b), density of states plot in Fig.~\ref{fig3bandB}(c) and quantum conductivity plot in Fig.~\ref{fig3bandB}(d). Hence, one can say that 7aGNRsH-B  acts like a metallic structure  due to the periodicity of B atoms in the armchair direction imposed by the periodic boundary conditions in the first-principles calculations~\cite{topsakal08}.

In Fig.~\ref{fig4bandV} (a), we have plotted a side view of the optimized structure of one carbon atom vacancy in 7aGNRsH (7aGNRsH-V) along with maximally localized wannier functions of the top of the valence band states. For 7aGNRsH-V, the Fermi energy, $\epsilon_F = 0$ also exits below the top of the valence band states which can be seen in the bandstructures calculation in  Fig.~\ref{fig4bandV}(b), density of states plot in Fig.~\ref{fig4bandV}(c) and quantum conductivity plot in Fig.~\ref{fig4bandV}(d). Hence, one can say that 7aGNRsH-V also acts like a metallic composition due to the periodicity of ring-pentagon structures that is imposed by periodic boundary conditions in the first-principles calculations~\cite{topsakal08}.

Pure intrinsic semiconductor has no applications unless we change intrinsic semiconducting materials into the n-type or  p-type semiconducting or completely transform into  metallic where these materials behave like electrically  conductive for optoelectronic device applications. Therefore, we have been interested in finding influence  of electrical conductivity against frequency on these n-type, p-type or metallic materials as well as influence of Berry curvatures when fermions move throughout the Brillouin zone. In solids, e.g., graphene, when charge carriers are transported in the Brillouin zones of reciprocal space in a piecewise continuous manner from -X point to $\Gamma$-point to X point with the application of built in intrinsic electric field, then these charge carriers acquires a Berry phase which is obtained by integrating Berry curvature, $\Omega_n$ as~\cite{zak89},
\begin{equation}\label{zak}
  \bm{\Omega}_n(\bm{k}) = \bm{\nabla_k} \bm{\times} \bm{A}_n(\bm{k}) =  -Im \left<\bm{\nabla_k} u_{n\bm{k}}| \times| \bm{\nabla_k} u_{n\bm{k}} \right>,
\end{equation}
where Berry connection,  $\bm{A}_n(\bm{k})$ is defined in terms of cell-periodic Bloch states $u_{n\bm{k}}=\exp\left(-i\bm{k\cdot r}\right)\left|\psi_{n\bm{k}}\right>$ as,
\begin{equation}\label{zak}
 \bm{A}_n(\bm{k}) =\left<u_{n\bm{k}}|i\bm{\nabla_k}|u_{n\bm{k}}\right>.
\end{equation}
For calculation of electrical  conductivity, one can use Kubo-Greenwood formula of electron in crystals in the independent particle approximation as,
\begin{widetext}
\begin{equation}\label{kubo}
  \sigma_{\alpha\beta}\left(\hbar\omega\right) = \frac{ie^2\hbar}{N_kV_c} \sum_{\bm{k}} \sum_{n,m} \frac{f_{m\bm{k}}-f_{n\bm{k}}}{\epsilon_{m,\bm{k}}-\epsilon_{n,\bm{k}}} \frac{\left< \psi_{\bm{n}k} \left|v_\alpha\right| \psi_{\bm{m}k} \right> \left< \psi_{\bm{m}k} \left|v_\alpha\right|\psi_{\bm{n}k}  \right>}{\epsilon_{m,\bm{k}}-\epsilon_{n,\bm{k}}-\left(\hbar \omega + i \eta\right)}.
\end{equation}
\end{widetext}
In Eq.~(\ref{kubo}), the indices $\alpha, ~\beta$ denote direction of cartesian coordinates, $V_c$ is the cell volume, $N_k$ is the number of k-points used for sampling the Brillouin zone and $f_{nk}$ is the Fermi-Dirac distribution function. Also, $\hbar\omega$ is the  frequency, and $\eta$ is an adjustable parameter. The off-diagonal matrix is written as
\begin{equation}\label{v}
  \left< \psi_{\bm{n}\bm k} \left|\bm{v}\right| \psi_{\bm{m}\bm k} \right> = -\frac{i}{\hbar}\left(\epsilon_{m\bm k}-\epsilon_{n \bm k}\right) \bm{A}_{nm}\left(\bm{k}\right),
\end{equation}
where $m\neq n$ and $\bm{A}_{nm}$ is the Berry connection matrix~\cite{blount62}. Substituting Eq.~(\ref{v}) in Eq.~(\ref{kubo}), the conductivity Eq.~(\ref{kubo}) becomes
\begin{widetext}
\begin{equation}\label{kubo-1}
  \sigma_{\alpha\beta}\left(\hbar\omega\right) =  \frac{ie^2}{\hbar N_kV_c} \sum_{\bm{k}} \sum_{n,m}\left(f_{m\bm{k}}-f_{n\bm{k}}\right) \frac{\epsilon_{m\bm{k}}-\epsilon_{n\bm{k}}}{\epsilon_{m\bm{k}}-\epsilon_{n\bm{k}}-\left(\hbar\omega+i\eta\right)} A_{nm,\alpha}\left(\bm k\right)  A_{mn,\beta}\left(\bm k\right).
\end{equation}
\end{widetext}
We can decompose Hermitian (dissipative) and anti-Hermitian (reactive) parts of Eq.~(\ref{kubo-1}) and write the Hermitian and anti-Hermitian conductivities as
\begin{widetext}
\begin{eqnarray}\label{kubo-H}
  \sigma^H_{\bm k, \alpha\beta}\left(\hbar\omega\right) =  -\frac{\pi e^2}{\hbar V_c} \sum_{n,m}\left(f_{m\bm{k}}-f_{n\bm{k}}\right)
  \left(\epsilon_{m\bm{k}}-\epsilon_{n\bm{k}}\right) A_{nm,\alpha}\left(\bm k\right)  A_{mn,\beta}\left(\bm k\right)
  \bar{\delta}\left(\epsilon_{m\bm{k}}-\epsilon_{n\bm{k}}-\hbar\omega\right),\label{kubo-H}\\
  \sigma^{AH}_{\bm k, \alpha\beta}\left(\hbar\omega\right) =  \frac{ie^2}{\hbar V_c} \sum_{n,m}\left(f_{m\bm{k}}-f_{n\bm{k}}\right)
  Re \left[\frac{\left(\epsilon_{m\bm{k}}-\epsilon_{n\bm{k}}\right)}{\epsilon_{m\bm{k}}-\epsilon_{n\bm{k}}-\left(\hbar\omega+i\eta\right)}\right]  A_{nm,\alpha}\left(\bm k\right)  A_{mn,\beta}\left(\bm k\right),\label{kubo-AH}
\end{eqnarray}
\end{widetext}
 where,
\begin{equation}\label{bar}
 \bar{\delta}\left(\epsilon\right) = Im \left[\frac{1}{\epsilon - i \eta} \right].
  \end{equation}
In Fig.~\ref{fig5BerryCurvature}, we plot the Berry curvatures of fermions in intrinsic 7aGNRsH, as well as metallic 7aGNRsH-B and 7aGNRsH-V throughout the Brillion zones in reciprocal space. For intrinsic 7aGNRsH where Fermi energy stays at the middle of the bandgap, it can be seen from the Berry curvature plot that fermions are distributed throughout the Brillion zones, whereas for mettalics  7aGNRsH-B and 7aGNRsH-V, it can be seen that a discontinuity of fermions exists slightly away from the $\Gamma$-point due to the presence of Fermi Energy below the top of the valence band. In Fig.~\ref{fig6HVBcond}, we  plot the electrical conductivity as a function of photon energy for 7aGNRsH, 7aGNRsH-B and 7aGNRsH-V. Here we  find that intrinsic 7aGNRsH is electrically  inactive but peaks in  conductivities for metallic 7aGNRsH-B and 7aGNRsH-V indicate that metallic 7aGNRsH-B and 7aGNRsH-V are electrically active in the wide range of spectrum that spans from IR ( 1.7eV-1.24eV) to visible (3.3 eV-1.7eV) to UV (larger than 3.3eV) energy spectrum regime.

When strain is implemented in the armchair graphene nanoribbons with  odd number of zigzag edges, then the width of the graphene nanoribbons either decreases or increases depending on which kind of strain is applied at the boundaries of the ribbons. Hence, large modification of bandgap is expected due to application of strain.   For example, strain induced by compressive stress reduces the width of the ribbons while tensile stress increases the width of the ribbons. Since the bandgap of graphene nanoribbons depends on the width of the ribbon, we expect the fine tuning of bandgaps with application of strain from semiconductor to metal or viceversa. In this paper, we are  interested in tuning the bandgap from a semiconducting to metallic and then bringing it back to the semiconductor with the application of strain engineering. Hence, we consider armchair graphene nanoribbons with  7 zigzag edges  and then apply the compressive stress at the armchair edge. The armchair ribbons is also allowed to elongate in the x-direction. The schematics of the strained armchair graphene nanoribbon is shown in Fig.~\ref{fig7Strain}. One may consider 9, 11, 14 and so on and so forth armchair graphene nanoribbons and then apply similar kind of compressive stress at the armchair edges for tuning the bandgaps. However in this case, the bandgaps opening at $\Gamma$-point is smaller than 7 armchair GNRs and thus one expects bandgap closing at lower values of strain. On the other hand, one may consider 5 and 3 armchair GNRs but the bandgap opening in such nanoribbons is larger than 7 armchair GNRs and hence one may need to apply large strain to close the bandgaps, that may be impractical for the experiment. It is also possible to consider  the even number of armchair edges of the GNRs and then apply the  strain   so that one can tune the properties of bandgaps from metal to semiconductor. We  plot the width of optimized structures of strained intrinsic 7aGNRsH in Fig.~\ref{fig8Berry} (a) and band structures in Fig.~\ref{fig8Berry} (b). Clearly, the width of the ribbon decreases as the value of strain increases that can be seen in Fig.~\ref{fig8Berry} (a). As a result, bandgaps of printine 7aGNRsH decreases because width of the ribbons moves towards the metallic 6 armchair graphene nanoribbons. But for very large values of strain, approximately at 18$\%$, we find that out-of-plane deformations provide relaxed shape graphene nanoribbons (e.g., Fig.~\ref{fig8Berry} (a), lower plot). As a result, fermions spread on a large surface area of armchair graphene nanoribbons that cause to increase the bandgap (see Fig.~\ref{fig8Berry}(b)). Berry curvature of fermions for pristine 7aGNRsH at several strain values is plotted in Fig~\ref{fig8Berry}(c). As can be seen, for unstrained 7aGNRsH, fermions are spread through out the Brillion zone in the reciprocal space but for strained 7aGNRsH that have out-of-plane deformations, fermions are localized near $\Gamma$-point in the reciprocal space of Brillion zone.

In Fig.~\ref{fig9Berry}(a), we have plotted the width of optimized relaxed structures of strained 7aGNRsH-B and band-structures in Fig.~\ref{fig9Berry}(b). Again, at about 18$\%$ of strain value, the 7aGNRsH-B has out-of-plane deformations that provide an enhancement of the bandgaps. Note that Fermi energy is set to be at 0 eV which is far below the top of the valence bands and these structures still viewed as metallic. The Berry curvatures for 7aGNRsH-B at several strain values through out the Brillion zone  are plotted in Fig.~\ref{fig9Berry}(c). Since Berry curvature corresponds to the gradient of all the filled  states up to the Fermi level but Fermi level stays below the top of the valence bands, one expects a  cut-off curvature away from the $\Gamma$-point that can also be reflected in Fig.~\ref{fig9Berry}(c). The cut-off curvature moves towards the $\Gamma$-point as we increase the strain (red plot of Fig.~\ref{fig9Berry}(c)) but again moves away from the $\Gamma$-point for $18\%$ strain (blue plot of Fig.~\ref{fig9Berry}(c)) due to having out-of-plane deformations.  In Fig.~\ref{fig10Berry}(a), we have plotted the width of optimized relaxed structures of strained 7aGNRsH-V, band structures in Fig.~\ref{fig10Berry}(b) and Berry curvatures in Fig.~\ref{fig10Berry}(c). Since 7aGNRsH-V is also metallic under several strain values that contains out-of-plane deformations regardless of strain conditions,  the cut-off in Berry curvature always moves away from $\Gamma$-point  because an increase in strain also increases  the out-of-plane deformations.

In Fig.~\ref{fig11Kubo}, we demonstrate the effect of strain on the electrical conductivity of armchair graphene nanoribbons. In Fig.~\ref{fig11Kubo} (a) for 7aGNRsH, it can be seen that unstrained 7aGNRsH is electrical inactive but becomes electrically active for strained cases in a wide range of energy spectra (e.g. energy of IR spectrum for strain up to 12$\%$ but energy of visible spectrum for strain of 18$\%$). Note that 18$\%$ of strain on 7aGNRsH  has out-of-plane deformations. On the other hand, for 7aGNRsH-B in Fig.~\ref{fig11Kubo} (b), peaks in electrical conductivity can always be seen in the IR (i.e., 1.24meV-1.7eV) and visible (i.e., 1.7eV-3.3eV) energy spectrum regimes because these materials are always metallic due to the presence of periodicity of B doping atom in 7aGNRsH-B. Also, for 7aGNRsH-V in Fig.~\ref{fig11Kubo} (c), peaks in electrical conductivity can always be seen in the IR (i.e., 1.24meV-1.7eV), visible (i.e., 1.7eV-3.3eV) and UV ( larger than 3.3eV) energy spectrum regimes  because these materials are always metallic due to the presence of periodicity of ring-pentagon in the crystal lattice.

Intrinsic semiconducting pristine unstrained 7aGNRsH is electrically non-conductive (e.g., see Fig.~\ref{fig6HVBcond}). The fundamental requirement for the development of advanced electronic device architectures, based on graphene nanoribbons, requires the modulation of the band structure and charge carrier concentration. The modulation can be achieved  by substituting specific carbon atoms in the hexagonal graphene lattice so that pristine 7aGNRsH  transforms into the  p- or n-type dopant heteroatoms~\cite{kawai15,cloke15}. Hence, we expect armchair graphene nanoribbons of dopant heteroatoms  become electrically  conductive.  For example, site-specific substitution of two boron doping of semiconducting armchair graphene nanoribbons, (e.g., see Fig.~\ref{fig12-2B-Band}(a (i))) turns into p-type nanoribbons, as can be seen in the band structures plot in Fig.~\ref{fig12-2B-Band}(b). The maximally localized wannier functions of the top of the valence band and impurity state above Fermi energy, $\epsilon_F=0$ are shown in Fig.~\ref{fig12-2B-Band}(a (ii,iii))).  In the band structures calculation of two boron doped  7aGNRsH in Fig.~\ref{fig12-2B-Band}(b), the band energies obtained from pw.x and wannier.90 method are in excellent agreement. Note that the optimized structure of site-specific substitution of 2 boron atom in pristine 7aGNRsH, as shown in Fig.~\ref{fig12-2B-Band}(a), is exactly the same structure that was experimentally synthesised in Ref.~\cite{cloke15}. The calculated bandgap between top of the valence band and dopant state is 0.6eV, which is 0.2eV smaller than the bandgap obtained from GW method reported in Refs.~\cite{cloke15}. On the other hand, the quasi-particle bandgap between the top of the valence band and second conduction band is 2.0eV, which is 0.5eV smaller than the bandgap obtained from GW method but 2.0 eV bandgap matches with the results obtained from LDA method (see references in~\cite{cloke15}).  As can be seen in the band structure calculation in Fig.~\ref{fig12-2B-Band}(b), the dopant state exists above the Fermi energy, $\epsilon_F = 0$ which shares an empty p-orbital state with an extended $\pi$-band dopant functionality that eventually modulate the band structure and charge carriers. Hence, non-vanishing electrical conductivity is expected in such p-type boron doped armchair graphene nanoribbons due to the presence of dopant states. The dopant states can also be seen in the DOS and quantum conductivity plots in Fig.~\ref{fig12-2B-Band}(c,d) at about 0.25eV.  Finally, in Fig.~\ref{fig12-2B-Band}(e,f), the goal is to provide first-principles calculation results for the Berry curvature and the electrical conductivity in such p-type boron doped armchair graphene nanoribbons~\cite{cloke15}. In contrast with the cut-off bands in Berry curvature for metallic 7aGNRsH-B(Fig.~\ref{fig5BerryCurvature}, red graph) and 7aGNRsH-V(Fig.~\ref{fig5BerryCurvature}, blue graph), the Berry curvature plot of 2 boron site substitution in 7aGNRsH in Fig.~\ref{fig12-2B-Band}(e) is continuous throughout the Brillion zone that has peaks near X-point. One can say that fermions are localized more near the X-point than the $\Gamma$-point in the reciprocal space of the Brillion zone. In Fig.~\ref{fig12-2B-Band}(e), the peaks in  conductivity are mostly found in the IR energy spectrum ($<1eV$) regime but small peaks can also be observed in the visible energy spectrum regime ($\approx 2 eV$).

\section{Conclusion}\label{conclusion}
In summary, we have used first-principles calculations to investigate the role of strain, doping and vacancy on the band structures, electrical conductivity and Berry curvature of fermions in armchair graphene nanoribbons with 7 zigzag edges.  The results for band structures calculations show that pristine 7aGNRsH is semiconducting (e.g., see  Fig.~\ref{fig2bandH}) but turns to be metallic for boron doped 7aGNRsH (e.g., see Fig.~\ref{fig3bandB}) and one carbon atom vacancy in 7aGNRsH (e.g., Fig.~\ref{fig4bandV}) due to the periodicity  of boron and ring-pentagon in the  crystal lattice. In Fig.~\ref{fig8Berry}, influence  of strain on the Berry curvature of 7aGNRsH shows that fermions are spread through out the Brillion zone in the reciprocal space but localized near the $\Gamma$-points that have out-of-plane deformations.  For metallic 7aGNRsH-B, the Berry curvature plot in Fig.~\ref{fig9Berry} shows that fermions are localized near the  $\Gamma$-point but far away from the $\Gamma$-point for 7aGNRsH-V that is shown in Fig.~\ref{fig10Berry}. The results for electrical conductivities in Fig.~\ref{fig6HVBcond} show that pristine intrinsic  7aGNRsH is electrically  inactive but turns to be active in a wide range of energy spectrum regime that spans from IR ( 1.7eV-1.24eV) to visible (3.3 eV-1.7eV) to UV (larger than 3.3eV) for single atom boron doping as well as vacancy of one carbon in 7aGNRsH. Also, vanishing electrical conductivity in unstrained intrinsic 7aGNRsH  turns to be electrically active with the application of strain in  the wide range of IR, visible and UV energy spectrum, that is shown in Fig.~\ref{fig11Kubo}(a). In Fig.~\ref{fig11Kubo}(b) for 7aGNRH-B and Fig.~\ref{fig11Kubo}(c) for 7aGNRsH-V, peaks in electrical conductivities are always observed regardless of strain conditions. Finally, in Fig.~\ref{fig12-2B-Band} for two atom boron doped p-type semiconducting 7aGNRsH that has no periodicity of boron atom in the crystal lattice,  insight for the electrical conductivity of experimentally synthesised materials in Ref.~\cite{cloke15} has been provided. The results show that large peaks in electrical conductivity can be seen in the IR energy spectrum regime but experimentally detectable small peaks can also be seen in the visible energy spectrum regime. The first-principles calculations results for electrical conductivities in several kinds of 2D materials presented in this paper may be useful for making sensor devices and interconnects in the IR, visible and UV energy spectrum regimes.

\begin{acknowledgments}
The first-principles calculations have been performed at the bartik supercomputer cluster, Northwest Missouri State University. Some calculations have been performed at xsede clusters (www.xsede.org) as an startup allocation and compute canada supercomputer (computecanada.ca). SP  acknowledges Northwest Missouri State University for providing financial support to present these results at American Physical Society March meeting at  Chicago (2022). RM is acknowledging the support of NSERC Discovery and CRC Programs.
\end{acknowledgments}

%

\end{document}